\documentclass[prx,aps,twocolumn,superscriptaddress,bibnotes]{revtex4}

\usepackage{dcolumn}
\usepackage{bm}
\usepackage{amssymb}
\usepackage{color}
\usepackage[pdftex]{graphicx}

\usepackage[mathscr]{euscript}

\usepackage{amsmath}
\usepackage[linesnumbered,ruled]{algorithm2e}

\usepackage{url}
\usepackage[section]{placeins}
\usepackage[colorlinks=true,linkcolor=blue,citecolor=blue]{hyperref}
\usepackage{comment}

\begin{document} 
\let\vec\mathbf

\title{Enhancement of Atomic Diffusion due to Electron Delocalization in Fluid Metals}

\author{Chen Cheng}
\thanks{ORCID:https://orcid.org/0000-0003-3224-1231}
\affiliation{Department of Physics, University of Virginia, Charlottesville, VA 22904, USA}

\author{Gia-Wei Chern}
\thanks{corresponding author. }
\affiliation{Department of Physics, University of Virginia, Charlottesville, VA 22904, USA}

\date{\today}

\begin{abstract}
We present a general theory of atomic self-diffusion in the vicinity of a Mott metal-insulator transition in fluid metals. Upon decreasing the electron correlation from the Mott insulating phase, the delocalization of electrons gives rise to an increasing attractive interatomic interaction, which is expected to introduce an additional friction, hence reducing the atomic diffusivity. Yet, our quantum molecular dynamics simulations find an intriguing enhancement of the diffusion coefficient induced by the emerging attractive force.  We show that this counterintuitive phenomenon results from the reduction of the repulsive core and the suppression of the attractive tail by thermal fluctuations. The proposed scenario is corroborated by the Chapman-Enskog theory and classical molecular dynamics simulations on a standard liquid model based on the Morse potential. Our work not only provides a general mechanism of the attraction-facilitated diffusion enhancement in simple liquids, but also sheds new lights on the nontrivial effects of electron correlation on atomic dynamics. 
\end{abstract}

\maketitle

The kinetic properties of liquids, such as self-diffusion, viscosity, and thermal conductivity, are a subject of both fundamental interest as well as technological importance~\cite{Frenkel1955,Balucani94,Liboff03}. In particular, the diffusivity is a fundamental property that encodes the information of how interatomic interactions affect the random atomic motion in a fluid. For example, the Einstein equation, $D =  k_B T / \zeta$, relates the atomic self-diffusion $D$ to the friction coefficient $\zeta$ which partially quantifies the effect of interatomic forces~\cite{Cussler2009,Zwanzig1983}; here the temperature $T$ underscores the stochastic nature of the thermally activated diffusive motion. Qualitatively, a stronger interatomic force is thus expected to increase the friction and viscosity, hence giving rise to a reduced atomic diffusion.

The interatomic potential in liquids generally consists of a short-distance sharp repulsion originating from the Pauli exclusion of overlapping electron orbitals and a longer-ranged attraction which provides the cohesive energy~\cite{Hansen06,Chaikin95}.  Phenomenologically, the interatomic forces in simple liquids consisting of, e.g. noble-gas atoms are well approximated by the familiar Lennard-Jones (LJ) potential~\cite{Jones31,Stone13}. Monatomic liquid metals such as alkali fluids comprise another important family of simple liquids~\cite{Scopigno2005,March90}. The interatomic interactions in liquid metals, however, are significantly more complicated due to the formation of metallic bonds~\cite{Fiolhais1995,Gonzalez1993,Canales1994,Wax2007,Lewis1977}.  Their effective potential exhibits a softer core~\cite{Kambayashi1994,Canales1997,Canales1999,Anento1999} compared with LJ-type potentials and a long-range oscillating tail, similar to the RKKY interaction in itinerant magnets~\cite{Kubler00}. 

While the repulsive core is mostly responsible for the the emergence of short-range order that is characteristic of a liquid state~\cite{Chandler70,Weeks71}, extensive works over past decades have shown that the repulsive force also dominates the dynamical properties, especially in the dense limit~\cite{Ascarelli68,Chandler74,Protopapas73}.  The addition of an attractive tail to the repulsion generally leads to further friction and a reduced self-diffusion coefficient~\cite{Helfand61,Davis67,Gaskell71,Straub92}. However, the intricate interplay between the repulsive and attractive interactions has yet to be studied in detail, especially in the context of liquid-state Mott metal-insulator transitions~\cite{Mott90}, where the localization of electrons results in a diminished attractive interatomic interaction. Such study is particularly important for understanding the metal-insulator transition in expanding alkali fluids along the liquid-vapor coexistence line~\cite{Hensel99}, where several experiments have suggested the important role played by the electron correlation~\cite{Hensel99,Freyland1979,Hanany1983,Warren1989}.

More generally, the effects of electron correlation on the atomic dynamics in a liquid metal remain a largely uncharted territory in the research of both liquid-state physics and correlated electron systems. This is partly due to the lack of proper molecular dynamics (MD) methods to study such effects. The widely used classical MD~\cite{Allen87} methods which reply on empirical interatomic potentials fail to describe electronic phase transitions. On the other hand, state-of-the-art quantum MD~\cite{Iftimie05,Marx09} simulations based on density functional theory cannot properly include the strong electron correlation effects, hence are inadequate for simulating the Mott transition.

In this paper, we present a comprehensive study for the effects of electron correlation on the self-diffusion coefficient of liquid metals with correlated electrons. We demonstrate and characterize an unusual maximum of atomic diffusion near the Mott transition by employing a new quantum MD scheme~\cite{Chern2017,Chern2019} based on the Gutzwiller/slave-boson method, which offers an efficient and qualitatively correct description of the Mott transition~\cite{Gutzwiller1963,Gutzwiller1964,Gutzwiller1965,Kotliar1986}. Intriguingly, we further show that a similar diffusion maximum also occurs in classical simple liquid models when an attractive interatomic potential is added to the short-range repulsive interaction, a process which mimics the emerging cohesive force due to electron delocalization.   A general theory is also developed to account for this attraction-facilitated enhancement of the atomic diffusion in simple liquids.


To investigate the effects of electron correlations on motion of atoms in a liquid, we consider the following minimum liquid-metal model, 
\begin{eqnarray}
	\label{eq:H0}
	{\mathscr{H}} =  \sum_i \frac{|\Vec{p}_i|^2}{2m} +  \frac{1}{2} \sum_{i \neq j} \phi(|\mathbf r_i - \mathbf r_j|) + \hat{\mathcal{H}}_e\bigl( \{ \mathbf r_i \} \bigr),
\end{eqnarray}
where $\mathbf r_i$ and $\mathbf p_i$ are the classical position and momentum variables, respectively, of the $i$-th atom. The first term is the classical kinetic energy, while the second term represents the short-range repulsive pair potential. The attractive cohesive forces are provided by itinerant electrons described by the Hamiltonian: 
\begin{eqnarray}
	\label{eq:hubbard}
	\hat{\mathcal{H}}_e\bigl( \{ \mathbf r_i \} \bigr) =  \sum_{ij, \sigma} h(|\mathbf r_i - \mathbf r_j|) \hat{c}^{\dagger}_{i, \sigma} \hat{c}^{\,}_{j, \sigma}
	+ U \sum_{i}  \hat{n}_{i\uparrow} \hat{n}_{i_\downarrow}  
\end{eqnarray}
which can be viewed as a generalization of the Hubbard model~\cite{Hubbard63} to an atomic liquid. Here the first term describes electron hopping between atoms with $\hat{c}^\dagger_{i, \sigma}$ ($\hat{c}^{\,}_{i, \sigma}$) being the creation (annihilation) operator of an electron of spin $\sigma = \uparrow$,~$\downarrow$ at the $i$-th atom, and $\hat{n}_{i,\sigma} = \hat{c}^\dagger_{i, \sigma} \hat{c}^{\,}_{i, \sigma}$ the corresponding number operator. The Hubbard parameter~$U$ in the second term encapsulates the on-site Coulomb repulsion of electrons. This Hubbard liquid model is based on the tight-binding MD formulation~\cite{khan89,wang89,goedecker94} with the inclusion of the Hubbard term.

It is worth noting that the above Hamiltonian in Eqs.~(\ref{eq:H0}) and (\ref{eq:hubbard}) also serves as a minimum model for metallic alkali fluids~\cite{Rose81,Chapman88,Yonezawa82}. 
Assuming well-localized Wannier functions, we employ a hopping function which decays exponentially: $h(r) = -t_0 \, e^{-r/ \xi}$, where $\xi$ and $t_0$ set the length and energy scales, respectively, of the model. On the other hand, a faster-decaying function $\phi(r) = \phi_0 \,\exp[ -  (r / \lambda) - b( r/\lambda)^4  ]$ is used for the much sharper repulsive core. In this work, a set of model parameters $\phi_0  = 4.17\, t_0$, $\lambda = 0.86\, \xi$, and $b = 0.1$ is used.

While several advanced techniques such as dynamical mean-field theory~\cite{Georges96,Kotliar06} can provide accurate description of the Mott transition physics, they are computationally too expensive for quantum MD simulations. To this end, we apply the recently developed quantum MD scheme based on the Gutzwiller/slave-boson method to simulate the above Hubbard liquid model~\cite{Chern2017,Chern2019}. In this approach, the collective electron behaviors such as the local double-occupation are encoded in the slave-boson degrees of freedom~\cite{Kotliar1986}, while the quasi-particles are described by a tight-binding Hamiltonian with renormalized hopping constants: $t^*_{ij} = \mathcal{R}_{i} \mathcal{R}_{j} \, h(r_{ij}) $, where $r_{ij} = |\mathbf r_i - \mathbf r_j|$. The renormalization factors $\mathcal{R}_i$ depend on the slave-boson amplitudes and need to be solved self-consistently with the spectrum of quasi-particles~\cite{Lanata2012,Sandri2013}.

\begin{figure}
\centering
\includegraphics[width=0.99\columnwidth]{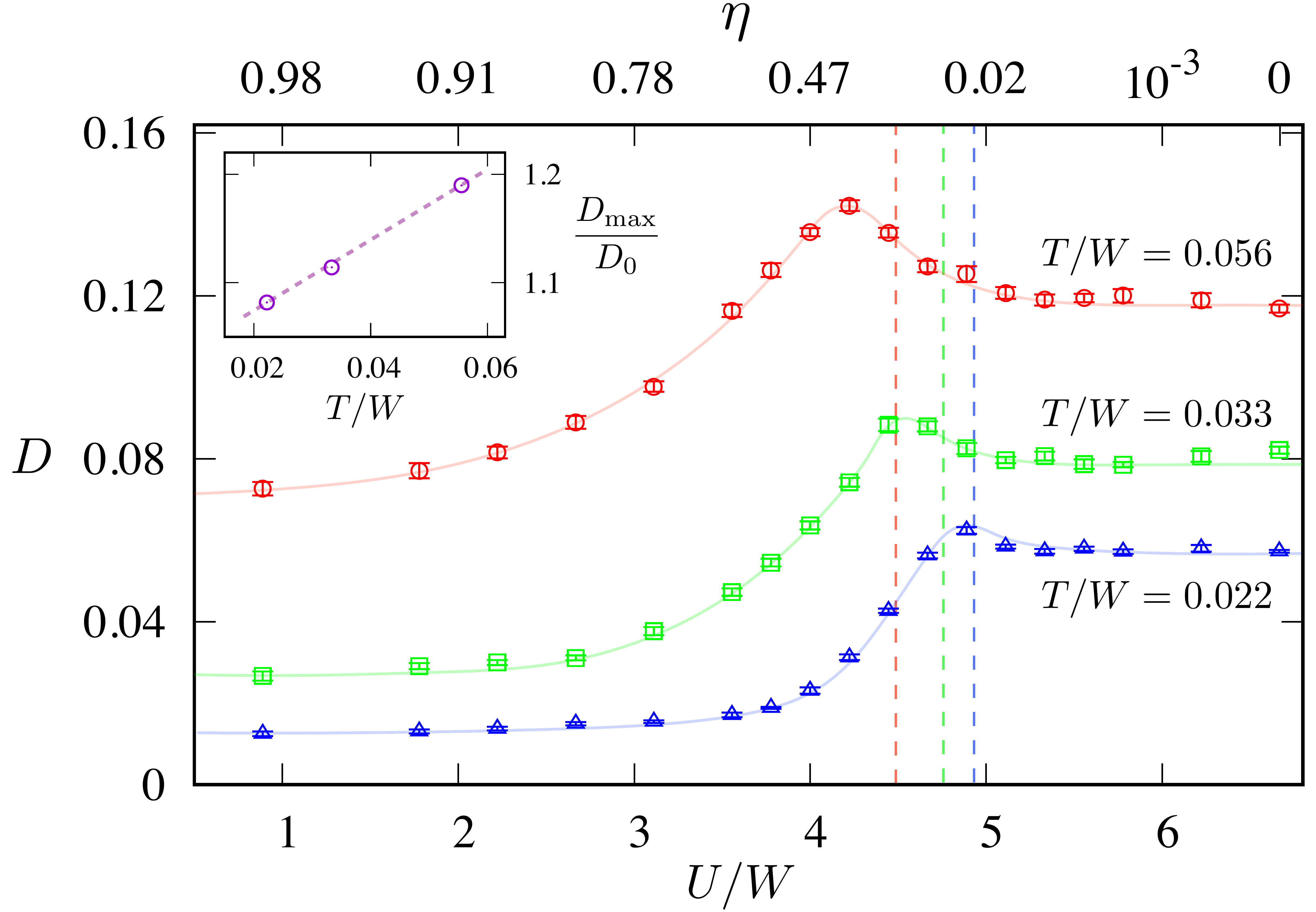}
\caption{The atomic self-diffusion coefficient $D$ versus the Hubbard parameter $U$ from the GMD simulations of the Hubbard liquid model Eq.~(\ref{eq:H0}) at three different temperatures. The dashed lines indicate estimate of the critical $U_c$ for the Mott transition. $W$ is the absolute value of electron energy at $U = 0$ at $T = 0.0042 t_0$, $W = 0.1875 t_0$. The inset shows the temperature dependence of $D_{\rm max} / D_{0}$, which provides a measure of the enhancement; here $D_{\rm max}$ is the maximum diffusion coefficient and $D_0$ is the value in the $\eta \to 0$ limit.}
\label{fig:GMD_diffusion}
\end{figure}

We perform the Gutzwiller MD (GMD) simulations on systems of $N = 50 \sim 200$ atoms, with periodic boundary conditions. The system volume is fixed by setting the Wigner-Seitz radius $r_s = (3V/4\pi N)^{1/3}$ to be $r_s = 1.9 \xi$. The temperature $T$ is controlled by the Langevin thermostat with a small damping. The atomic forces, computed using the Hellmann-Feynman formula $\mathbf F_i = -\langle \partial {\mathscr{H}}  / \partial \mathbf r_i  \rangle$, can also be derived from an effective pair-like potential $\mathbf F_i = - \partial V(r_{ij}) / \partial \mathbf r_i$. It consists of two parts:
\begin{eqnarray}
	\label{eq:V_eff}
	V(r_{ij}) =   \phi(  r_{ij})   -2   \bigl( \langle \hat{c}^\dagger_i \hat{c}^{\,}_j \rangle + \langle \hat{c}^\dagger_j \hat{c}^{\,}_i \rangle \bigr)  h(r_{ij}) ,
\end{eqnarray}
where the factor 2 accounts for the spin degeneracy, and $\langle \hat{A} \rangle = {\rm Tr}(\hat{\varrho}_G \hat{A} )$, with $\hat{\varrho}_G$ being the many-body electron density matrix within the Gutzwiller approximation, denotes the quantum average of operator $\hat{A}$. The first term describes the short-range repulsive core, while the second term denotes a renormalized attractive force. It should be noted the forces here are not exactly pair-wise since the electron reduced density matrix $ \langle \hat{c}^\dagger_i \hat{c}^{\,}_j \rangle$ depends also on atoms in the neighborhood of bond $(ij)$. Given the atomic forces, a velocity-Verlet algorithm is used to integrate the Langevin equation, and the self-diffusion coefficient is computed from the velocity autocorrelation:~\cite{Allen87,Rapaport95} 
$
	D = \frac{1}{3N} \sum_{i=1}^N \int_0^{\infty} \langle \mathbf v_i(0) \cdot \mathbf v_i(t) \rangle dt.
$

Fig.~\ref{fig:GMD_diffusion} shows the atomic self-diffusion coefficient $D$ as a function of Hubbard repulsion $U$  obtained from the GMD simulations for three different temperatures. The critical $U_c$ of the Mott transition, estimated by the vanishing of the double-occupancy~\cite{supp1}, are indicated by the vertical lines. Perhaps the most remarkable feature is the non-monotonic behavior of the diffusion coefficient which was first reported in Ref.~\cite{Chern2017}. The diffusion behavior at small $U$ can be understood within the framework of Stokes-Einstein relation. Assuming that atomic dynamics in this regime is dominated by the electron-mediated attraction, the increase of self-diffusion coefficient thus comes from a weakened attractive force. 
As indicated in Eq.~(\ref{eq:V_eff}), the effective attractive interaction relies on the delocalization of electrons, we thus introduce the following renormalization factor to characterize the weakening of the attractive force:
$	\eta \equiv  \langle \hat{c}^\dagger_i \hat{c}^{\,}_j \rangle  / \langle \hat{c}^\dagger_i \hat{c}^{\,}_j \rangle_{U=0} , $
where $\langle \cdots \rangle$ denotes both quantum average as well as spatial and ensemble averages from MD simulations. Within the Gutzwiller approximation, this factor is approximated as $\eta \sim \langle \mathcal{R}_i \mathcal{R}_j \rangle $.
The top horizontal axis of Fig.~\ref{fig:GMD_diffusion} shows the $\eta$ at $T/W = 0.056$, increasing the Hubbard~$U$ leads to a reduced $\eta$ and a weakened attractive force, which in turn increases the atomic diffusion. As the Hubbard parameter is greater than the critical $U_c$, the nearly complete localization of electrons leads to a pure repulsive interatomic interaction. The resultant diffusion coefficient is thus independent of $U$ in this Mott insulating phase, giving rise to the leveled curves on the right side of Fig.~\ref{fig:GMD_diffusion}.

\begin{figure}
\centering
\includegraphics[width=0.99\columnwidth]{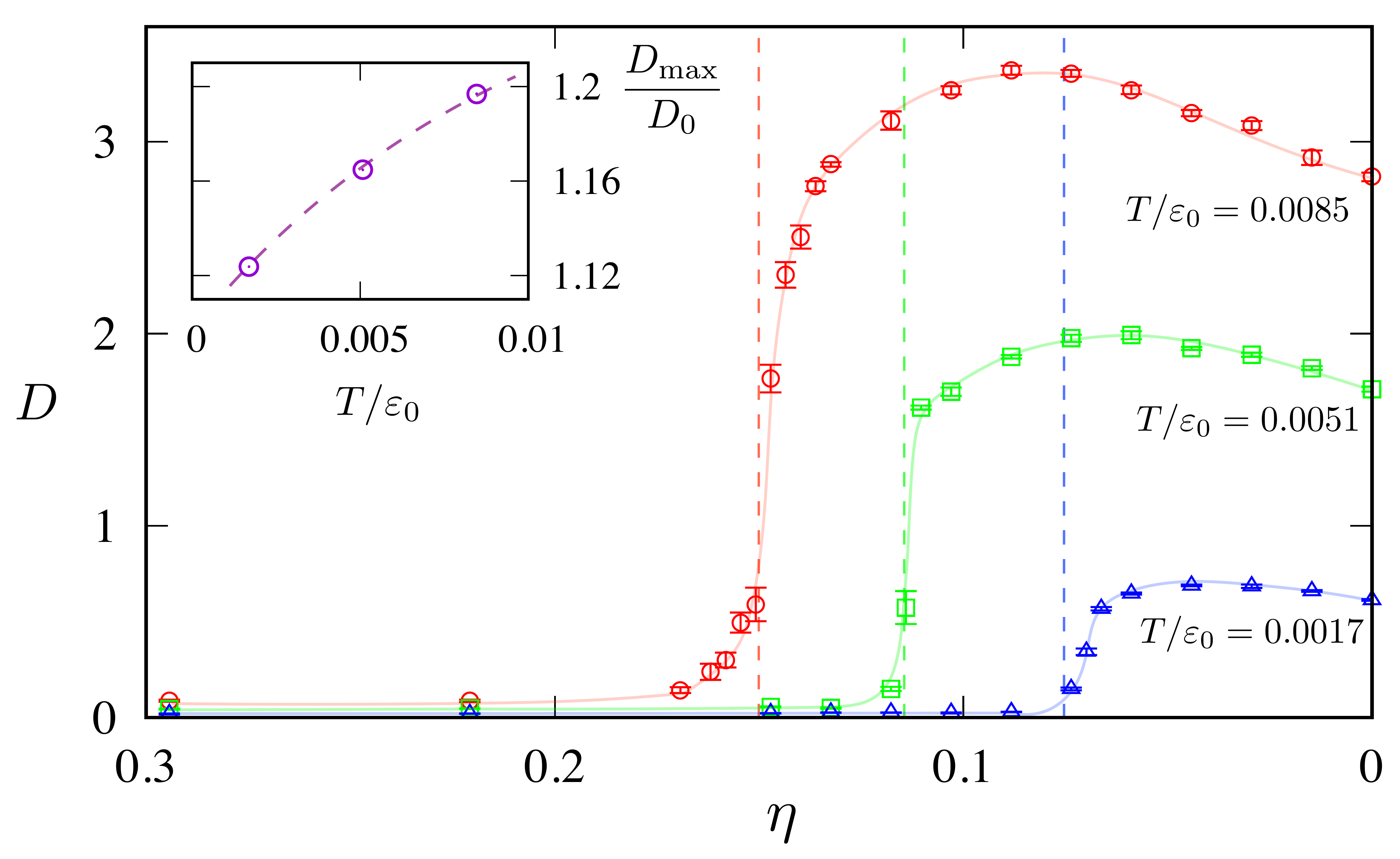}
\caption{The self-diffusion coefficient $D$ versus the renormalization parameter $\eta$ obtained from classical MD simulation of the generalized Morse potential Eq.~(\ref{eq:morse}). The dashed lines indicate the liquid-gas transitions at the corresponding temperatures. The inset shows the enhancement of self-diffusion defined as $D_{\rm max}/ D_0$, where $D_{\rm max}$ is the maximum diffusion coefficient and $D_0$ is the value at $\eta = 0$.}
\label{fig:CMD_diffusion}
\end{figure}

The above scenario based on a weakened attraction, however, could not explain the maximum of~$D$ in the vicinity of the Mott transition. In particular, due to the opposite signs of attractive and repulsive forces, their interplay near $U_c$ could lead to interesting dynamical behaviors. To this end, we consider an effective interatomic pair potential which can be viewed as a generalization of the well-known Morse potential~\cite{Varshni1957,Morse1929}
\begin{eqnarray}
	\label{eq:morse}
	V(r) = \varepsilon_0 \bigl( e^{-r/\ell_{+}} - \eta \, e^{-r / \ell_{-}} \bigr),
\end{eqnarray}
where $\varepsilon_0$ sets the energy scale, $\ell_{\pm}$ denote the ranges of the repulsive/attractive interactions. Importantly, we introduce a dimensionless parameter $\eta$ to mimic the renormalization of the attractive forces discussed above. The range of the repulsive core is in general smaller than that of attraction, for simplicity we set $\ell \equiv \ell_- = 2 \ell_+$. With this ratio, the special point $\eta = 2/e$ corresponds to the Morse potential $V(r) = \varepsilon_0 \{1-\exp[-(r-\ell)/\ell] \}^2$.

\begin{figure}
\centering
\includegraphics[width=0.99\columnwidth]{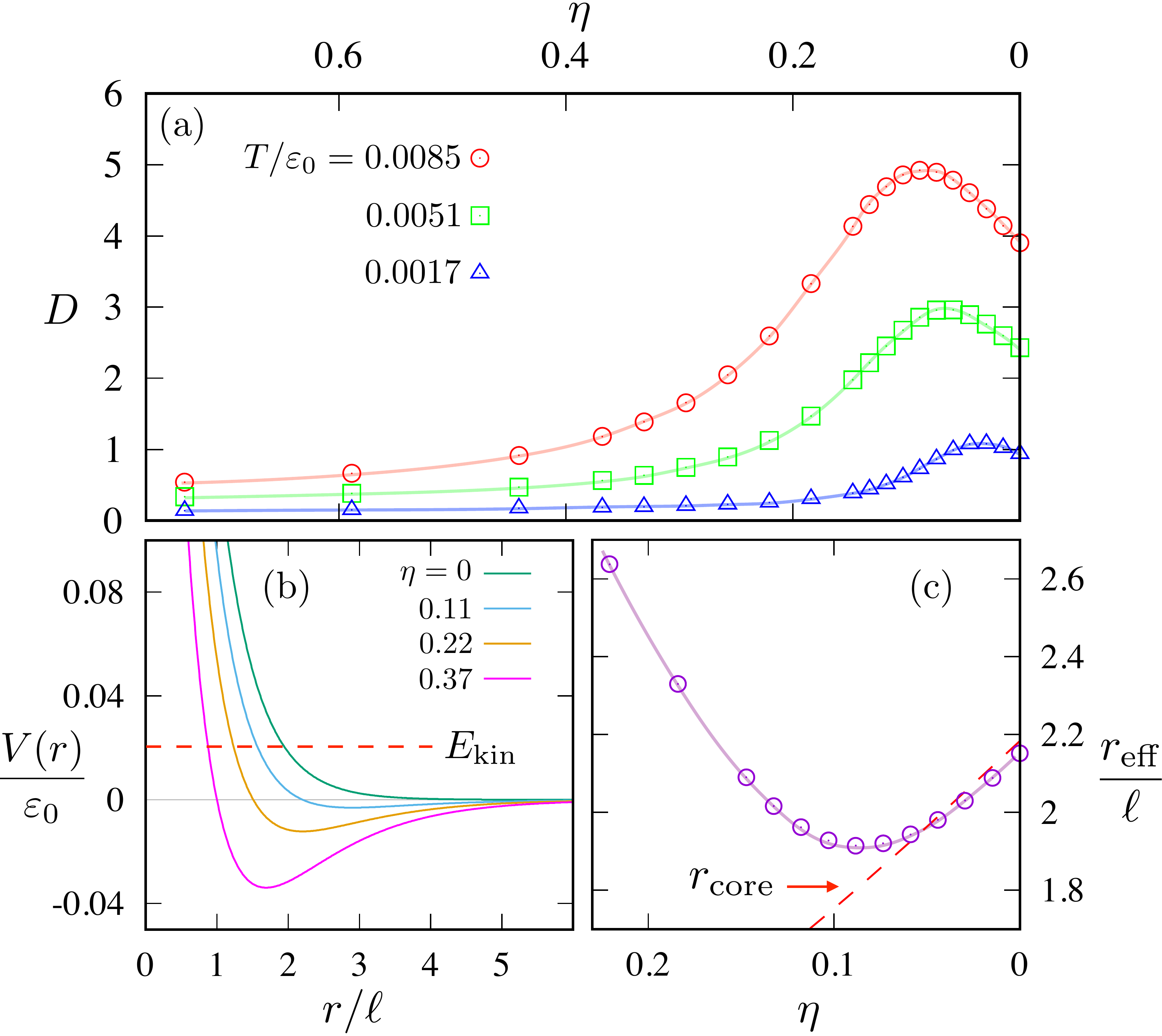}
\caption{(a) The diffusion coefficient of the generalized Morse potential Eq.~(\ref{eq:morse}), shown in panel~(b), versus the weakening factor $\eta$ computed using the Chapman-Enskog theory. (c)~Effective radius $r_{\rm eff}$ of the scattering cross-section versus $\eta$ for $T / \varepsilon_0 = 0.0085$. The red dashed line shows the radius of repulsive core determined by condition $V(r_{\rm core}) = E_{\rm kin} = 3 k_B T/2$, corresponding to the dashed line in panel~(b). }
\label{fig:Chapman}
\end{figure}

The generalized Morse potential is studied using standard classical MD simulations in the $NVT$ ensemble with up to $N = 1000$ atoms. We consider a relatively dilute system with an~$r_s = 4.5 \ell$. Fig.~\ref{fig:CMD_diffusion} shows the self-diffusion coefficient versus the renormalization parameter $\eta$ for three different temperatures. At large $\eta$, the strong attraction binds the atoms into a liquid state, it has a much smaller diffusion coefficient due to its high density. Upon reducing the parameter~$\eta$, the system undergoes a first-order liquid-gas transition, as indicated by the vertical lines in Fig.~\ref{fig:CMD_diffusion}. More relevant to our main interest here is the non-monotonic behavior of the diffusion coefficient as $\eta$ is further reduced in the gas phase. Importantly, this result which is similar to that of the Hubbard liquid model indicates that the mechanism of the diffusion maximum can be understood from this relatively simple classical liquid model.

To this end, we employ the Chapman-Enskog theory~\cite{Chapman70}, which provides an accurate description for dilute liquids or gases based on binary collisions, to study the kinetic properties of the generalized Morse potential. The diffusion coefficient is given by $D = \frac{3}{8} \left( \frac{\pi k_B T}{m}\right)^{1/2} \frac{1}{\pi  r_{\rm eff}^2 \,\rho }$, where $\rho = N/V$ is the atomic number density, $r_{\rm eff} = (m/\pi k_B T)^{1/4} \Omega_{(1,1)}^{1/2}$ is the effective radius of the scattering cross-section, and $\Omega_{(1,1)}$ is the collision integral~\cite{Chapman70}. As shown in Fig.~\ref{fig:Chapman}(a), the $D$ versus $\eta$ curves exhibit a clear maximum, which becomes more prominent with increasing temperature, consistent with the classical MD simulations in Fig.~\ref{fig:CMD_diffusion}. We note that the condensation phenomena in the strong attraction region, however, is beyond the Chapman-Enskog theory.

It is worth noting that this non-monotonic behavior is unexpected since even a small attractive interaction, which operates at a longer range, immediately introduces an additional friction, which in turn results in a reduced atomic diffusion.  While the enlarged diffusion coefficient is obviously due to a smaller $r_{\rm eff}$, we find that the intriguing reduction of the effective radius is well captured by a shrinking repulsive core. To see this, we define an effective core radius $r_{\rm core}$, which is similar to the so-called Boltzmann's hard-sphere diameter~\cite{Speedy89}, as the distance at which the potential energy equals the average kinetic energy of atoms, i.e. $V(r_{\rm core}) = E_{\rm kin} = 3k_B T /2$; see Fig.~\ref{fig:Chapman}(b).  As shown by the dashed lines in Fig.~\ref{fig:Chapman}(c), the $\eta$ dependence of the effective radius $r_{\rm eff}$ is well approximated by $r_{\rm core}$ almost all the way down to the minimum. 

\begin{figure}
\centering
\includegraphics[width=0.99\columnwidth]{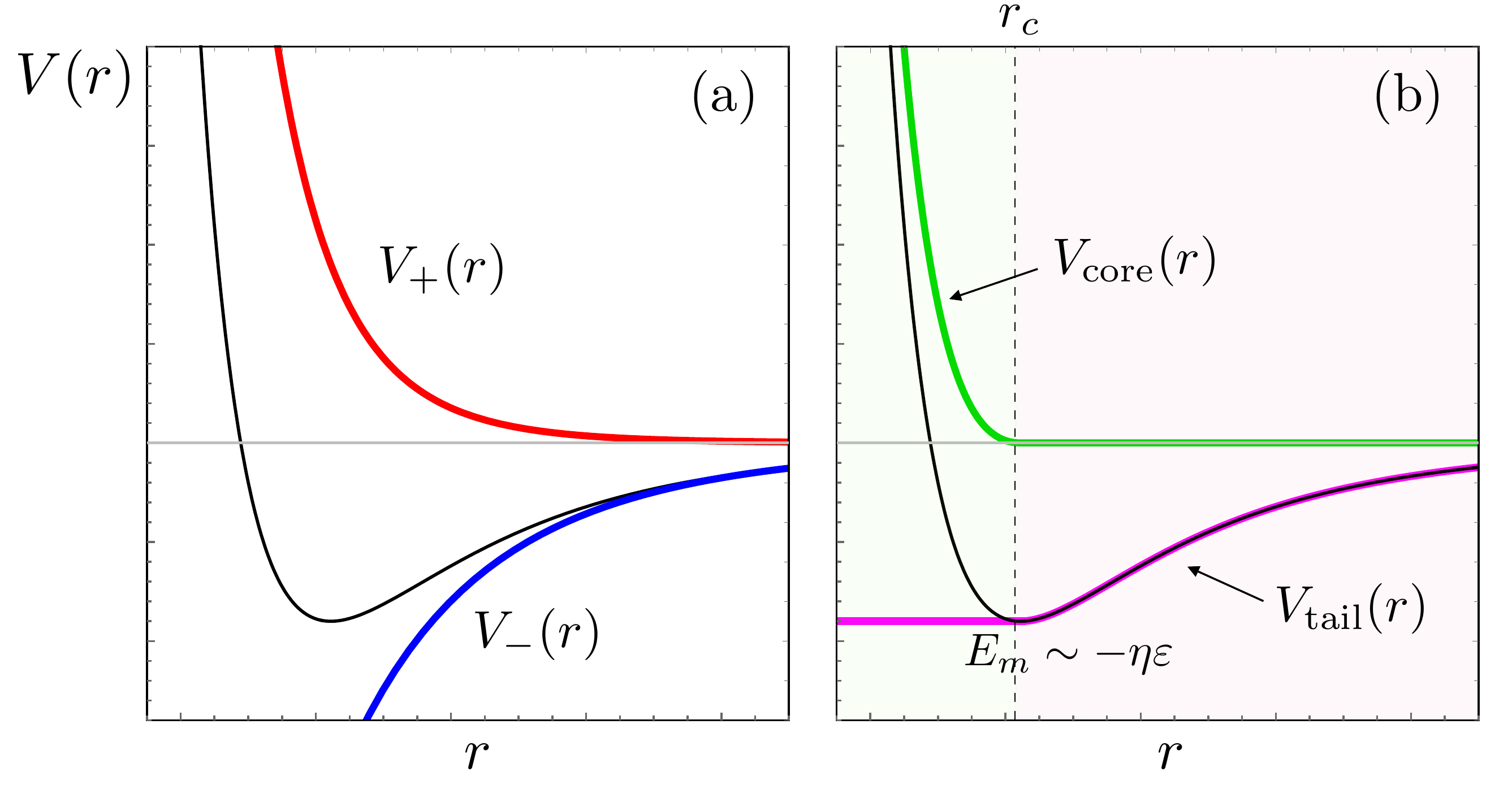}
\caption{(a) Decomposition of an interatomic potential $V(r)$ into its physical repulsive $V_+(r)$ and attractive $V_-(r)$ components. (b)~The same interaction potential can also be spatially separated into a repulsive core $V_{\rm core}(r)$ for $r < r_c$, and an attractive tail $V_{\rm tail}(r)$ for $r > r_c$. }
\label{fig:potential}
\end{figure}

Based on this observation, a theory of the diffusion maximum is presented in the following. As discussed in the introduction, an interatomic potential in general consists of a short-distance repulsive interaction $V_+$ and a longer-ranged attraction $V_-$; see Fig.~\ref{fig:potential}(a). Physically, these two components $V_\pm$ are of rather distinct origins. On the other hand, the importance of the inner repulsive core in determining the short-range correlation suggests a different decomposition $V = V_{\rm core} + V_{\rm tail}$~\cite{Weeks71} as shown in Fig.~\ref{fig:potential}(b). The force is purely repulsive in the core region, and is entirely attractive outside the core.  Based on the Einstein relation, the self-diffusion coefficient can then be expressed as $D = k_B T / (\zeta_{\rm core} + \zeta_{\rm tail})$, where the friction coefficients $\zeta$ are given by the force auto-correlation function. Here $\zeta_{\rm core}$ denotes the friction due to the repulsive core, while the effects of the attractive tail (including the cross terms) are subsumed into $\zeta_{\rm tail}$~\cite{Helfand61,Davis67,Gaskell71,Straub92}.

Since the atomic force is purely repulsive in the core region, its dynamical effect can be well approximated by an effective hard-sphere (HS) liquid system. Let $\sigma=\sigma(\eta)$ be the effective HS diameter for the repulsive core and replace $\zeta_{\rm core}$ by Enskog's formula for $\zeta_{\rm HS}(\sigma)$, the diffusion coefficient can then be expressed as
\begin{eqnarray}
	D = \frac{3}{8 \rho \pi [\sigma(\eta)]^2} \sqrt{\frac{\pi k_B T}{m}} \left[ 1 + R\!\left(\frac{|E_m|}{ k_B T} \right) \right]^{-1}.
\end{eqnarray}	
Here $R \equiv \zeta_{\rm tail} / \zeta_{\rm core}$ is the ratio of the two friction coefficients. Importantly, since $R$ is dimensionless, it depends on the energy minimum $E_m$ of the potential $V(r)$ only through the ratio $|E_m| / k_B T$~\cite{Chapman70,Straub92}. Next we consider the behavior of $D$ at small attraction. First, the effective HS diameter can be expanded as $\sigma(\eta) = \sigma_0 (1 - \alpha \eta) + \mathcal{O}(\eta^2)$, where the first-order coefficient $\alpha =  {\partial \ln \sigma}/{\partial \eta} > 0$ characterizes the compressibility, or softness, of the physical repulsive potential $V_+$; a larger $\alpha$ indicates a more significant attraction-induced core shrinkage. Since the energy minimum is due to the physical attractive interaction $V_-$, we have $E_m \sim -\eta \varepsilon$ for $\eta \ll 1$, where $\varepsilon$ is a characteristic energy scale, e.g. the binding energy, of the interatomic potential.  Finally, as $\zeta_{\rm tail}$ is also caused by the attractive potential $V_-$, the leading-order approximation for $R$ is thus $R = \Gamma \eta \varepsilon / k_B T + \mathcal{O}(\eta^2 \varepsilon^2 / T^2)$, where the coefficient $\Gamma$ depends on details of  both $V_\pm$ potentials, and is weakly dependent on temperature through, e.g. the effective HS diameter $\sigma_0$. It is worth noting that this leading dependence on $\varepsilon/k_B T$ has been demonstrated for the analytically solvable cases of the Sutherland~\cite{Chapman70} and square-well potentials~\cite{Korlipara89}.   The self-diffusion coefficient at small~$\eta$ thus becomes
\begin{eqnarray}
	D = \frac{D_0(T)}{ 1 - 2 \alpha \eta (T - T_{\rm th}) / T + \mathcal{O}(\eta^2) },
\end{eqnarray}
where $D_0$ is the diffusion coefficient of a pure repulsive~$V_+$, i.e. at $\eta = 0$, and $T_{\rm th} = \Gamma \varepsilon / 2 k_B \alpha$ is a threshold temperature. This expression encapsulates the two mechanisms of opposite effects due to a physically attractive interaction $V_-(r)$: the increase of atomic friction versus the reduction of the effective repulsive core. 

Crucially, the enhancement of the self-diffusion coefficient occurs only when temperature is above the threshold $T_{\rm th}$ when the effect of the longer-ranged attractive tail is suppressed by thermal fluctuations, consistent with both our quantum and classical MD simulations of the Hubbard and Morse liquid models, respectively.  The threshold temperature is proportional to the energy scale of the attractive tail $\Gamma \varepsilon$, and inversely proportional to the softness $\alpha$ of the repulsive potential $V_+$. As the strength of the attractive interaction is further increased, the higher order terms $\mathcal{O}(\eta^2)$ takes over, thus giving rise to the observed non-monotonic behavior. 

To summarize, we have presented a comprehensive theory for the intriguing phenomenon of attraction facilitated enhancement of atomic diffusion in simple liquids. 
While this phenomenon is illustrated by the Chapman-Enskog method in the dilute limit, our analysis shows the core shrinkage effect is also valid in dense liquids.
Our work highlights interesting kinetic phenomena that results from nontrivial interplay between repulsive core and the longer-ranged attractive interaction. 
In particular, our theory naturally explains a maximum atomic diffusivity in the vicinity of a Mott metal-insulator transition in liquid metals, a phenomenon which is demonstrated by quantum molecular dynamics simulations on correlated-electron liquid models. 

In addition to their fundamental importance, liquid metals such as sodium and lithium also find several technological applications as in nuclear reactors or concentrated solar power technologies~\cite{Baharoon2015, Pacio2013, Lorenzin2016} that operate at high temperatures. Several studies have suggested the importance role of electron correlation in the metal-insulator transition of alkali fluids~\cite{Freyland1979,Hanany1983,Warren1989}. As discussed above, the Hubbard liquid model offers a minimum theoretical description for these alkali fluids.   The numerical techniques and theoretical framework developed in this work pave the way toward a full understanding of the interplay between electron correlation and the atomic dynamics in correlated liquid metals.

 \bigskip

{\em Acknowledgements}.
We thank Gabriel Kotliar and Zhijie Fan for useful discussions and collaborations on related projects. This work is partially supported by the Center for Materials Theory as a part of the Computational Materials Science (CMS) program, funded by the U.S. Department of Energy, Office of Science, Basic Energy Sciences, Materials Sciences and Engineering Division. The authors also acknowledge the support of Research Computing at the University of Virginia.

\end{document}